\documentclass[aps,pre,twocolumn,showpacs,epsfig,floats]{revtex4}
\usepackage{amsmath}
\usepackage{color}
\usepackage{graphicx}
\usepackage{epsfig}
\usepackage{amsfonts}
\usepackage{mathrsfs}
\usepackage{amsmath}

\begin{document} 

\title{Drive Induced Delocalization in Aubry-Andr\'e Model}
 
\author{S. Ray, A. Ghosh and S. Sinha}
\affiliation{Indian Institute of Science Education and
Research-Kolkata, Mohanpur, Nadia-741246, India}
 
\date{\today}

\begin{abstract}
Motivated by the recent experiment by Bordia {\it et al} [{\it Nat. Phys.} {\bf 13}, 460 (2017)], we study single particle delocalization phenomena of Aubry-Andr\'e (AA) model subjected to periodic drives. In two distinct cases we construct an equivalent classical description to illustrate that the drive induced delocalization phenomena stems from an instability and onset of chaos in the underlying dynamics. 
In the first case we analyze the delocalization and the thermalization in a time modulated AA potential with respect to driving frequency and demonstrate that there exists a threshold value of the amplitude of the drive.
In the next example, we show that the periodic modulation of the hopping amplitude leads to an unusual effect on delocalization with a non-monotonic dependence on the driving frequency. 
Within a window of such driving frequency a delocalized Floquet band with mobility edge appears, exhibiting multifractality in the spectrum as well as in the Floquet eigenfunctions. 
Finally, we explore the effect of interaction and discuss how the results of the present analysis can be tested experimentally.

\end{abstract}


\maketitle

{\it Introduction:} Periodically driven quantum systems have been extensively used to study 
various phenomena like parametric resonance, quantum chaos, 
topological phases, etc \cite{par_reso,q_chaos,topo,anatoli_rev,dalibard_prx}.
Experimental and theoretical studies have gained interests in the context of
many body systems exhibiting thermalization in the presence of a periodic drive \cite{exp_drive,bloch,theo_drive,theo_drive1,nandkishore}.
On the other hand the fate of a driven many body localized (MBL) state is an emerging issue \cite{huse,adas} as has been demonstrated in a seminal experiment on ultra-cold atomic systems showing delocalization from the MBL phase \cite{bloch}.
It is a pertinent question to ask how generic is this phenomenon and whether there is any underlying principle
that explains drive induced delocalization of MBL phase.

To explain thermalization of a closed quantum system, the eigenstate thermalization hypothesis (ETH) has been proposed \cite{eth} and its close connection with the universal spectral properties of random matrix theory (RMT) has been explored \cite{eth_rmt,rigol}.
Similarly, spectral properties of a quantum system play a crucial role to characterize MBL phase and its transition to delocalized phase \cite{mbl_rmt,lev_stat}.
The delocalization of the MBL phase can be experimentally inferred from the decay of synthetically prepared excited states \cite{bloch_mbl} and signature of such phenomenon is expected to be captured from the spectral analysis; for a localized phase the level spacing distribution follows Poisson statistics whereas it changes to `Wigner Surmise' in the thermalized regime \cite{lev_stat}.
In an alternative approach, the route to thermalization can be explained by an underlying classical dynamics and has been extensively studied for Dicke model \cite{dicke,kdm}. 
Moreover, both the thermalization and the drive induced delocalization can involve departure from integrability associated with entropy generation. 
Hence it is natural to search for a classical route to the delocalization from MBL phase where periodic drive can trigger a dynamical instability and chaos. 

In the present work we consider an experimentally realizable model describing a system of bosons in the presence of a quasiperiodic potential and subjected to two different types periodic drives leading to distinct physical phenomena.
We start by constructing the corresponding classical Hamiltonians; the signature of delocalization obtained from the classical dynamics are compared with the entanglement entropy and the spectral properties of the Floquet operators of the system.
In the first model we consider the periodically modulated quasiperiodic potential analogous to the recent experimental scenario \cite{bloch}. We find that apart from the driving frequency, the strength of the periodic perturbation plays an equally important role in the thermalization of MBL state.  
In the second model we introduce a time periodic gauge field and counter-intuitively observe delocalized Floquet bands for a certain range of the driving frequency.
This delocalized band is separated from the localized states by a mobility edge and exhibits multifractal behavior indicating non-trivial correlations in the Floquet matrix \cite{kravtsov}. In this case the spectral analysis will be insufficient to characterize the onset of delocalization \cite{tgeisel}.
In the presence of interaction a similar scenario of delocalization is observed within a frequency interval associated with the level repulsion in the corresponding Floquet energy spectrum.

The most general Hamiltonian describing a system under periodic perturbation is given by,
\begin{equation}
H(t) = H_{0} + H_{1}(t)
\label{ham_t}
\end{equation}
where $H_{0}$ is time independent part and the time dependent part satisfies
$H_{1}(t + T) = H_{1}(t)$, where $T= 2\pi/\omega$ is the time period of the drive. 
The generic unitary time evolution of such system under the Hamiltonian $\hat{H}(t)$ in Eq.\ \ref{ham_t}
can be described by the Floquet operator $\hat{\mathcal{F}} = \hat{\mathcal{T}} e^{-i \int _{0}^{T} \hat{H}(t) dt}$, $\hat{\mathcal{T}}$ being the time ordering operator. We note that due to unitarity of $\mathcal{\hat{F}}$, the corresponding eigenvalue equation can be written as : $\mathcal{\hat{F}}|\psi_{\nu}\rangle = e^{-i \phi_{\nu}} |\psi_{\nu}\rangle$, where $\phi_{\nu}$ and $|\psi_{\nu}\rangle$ are the eigenphase and the eigenstate corresponding to the $\nu$th eigenmode of $\hat{\mathcal{F}}$. These single particle Floquet states can be decomposed in the real space as : $|\psi_{\nu}\rangle = \sum_{l} \psi _{\nu}(l) |l \rangle$, where $|l\rangle$ is the Wannier state and $\psi_{\nu}(l)$ is the amplitude of the Floquet state at  the $l$th lattice site. We will analyze the properties of these Floquet eigenmodes corresponding to the following driven systems.    
\\
{\it Model I:} We consider a periodically driven system of bosons within tight binding approximation given by the Hamiltonian; 
\begin{subequations}
\begin{eqnarray}
\hat{H}_0 &=& -J\sum_{l} \left(\hat{b}_{l}^{\dagger}\hat{b}_{l+1}+h.c.\right)  \\
\hat{H}_{1}(t) &=& \lambda \left(1+\epsilon f(t)\right) \sum_{l}\cos(2\pi \beta l)\hat{n}_{l}
\end{eqnarray}
\label{ham1}
\end{subequations}
where, $\hat{b}_{l}^{\dagger}$ and $\hat{n}_{l} = \hat{b}_{l}^{\dagger} \hat{b}_{l} $
are the creation and the density operator of the bosons at the $l$th lattice site respectively. $J$ is the hopping strength, $\lambda$ and $\beta$ denote the strength and incommensurability of the
quasiperiodic potential respectively. 
The drive is characterized by two parameters, frequency $\omega$ and the strength of the modulation $\epsilon$ which lies in the range $[0,1]$.
For simplicity we consider the form of the periodic modulation function in the interval $[0,2\pi]$, 
$f(x) = \theta (x - \pi) - \theta (\pi - x)$, where $\theta(x)$ is Heaviside step function.
In the absence of periodic drive $(\epsilon = 0)$, the Hamiltonian in Eq.\ \ref{ham1} represents the well known Aubry-Andr\'e model (AA) model \cite{aa_paper} which undergoes a localization transition for $\lambda > 2J$ when $\beta$ is chosen to be an irrational Diophantine number \cite{aa_paper,hanggi}.
In the rest of the paper we set $\beta = (\sqrt{5}-1)/2$, $\hbar = 1$, scale all the energies in the unit of $J$ (e.g. $\lambda \equiv \lambda/J$) and hence time in the unit of $1/J$. 

We first analyze the classical counterpart of single particle Hamiltonian in Eq.\ \ref{ham1}. By noting that the lattice translation operator is given by $\exp (ia\hat{p}/\hbar)$ where $a$ is the lattice constant and $\hat{p}$ is the momentum operator, an equivalent Hamiltonian corresponding to Eq.\ \ref{ham1} can be written as \cite{leboef},
\begin{equation}
H(X,P) = -2\cos P + \lambda \cos (2\pi \beta X)
\end{equation} 
where the scaled canonically conjugate variables satisfy the commutation relation $[X,P]=i$. The classical equation of motion obtained thereby are,
\begin{subequations}
\begin{eqnarray}
\dot{X} &=& 2\sin P \\
\dot{P} &=& 2\pi \beta \lambda (1 + \epsilon f(\omega t)) \sin (2\pi \beta X)
\end{eqnarray} 
\label{eq_of_motion}
\end{subequations}

\begin{figure}[ht]
\centering
\includegraphics[scale=0.115]{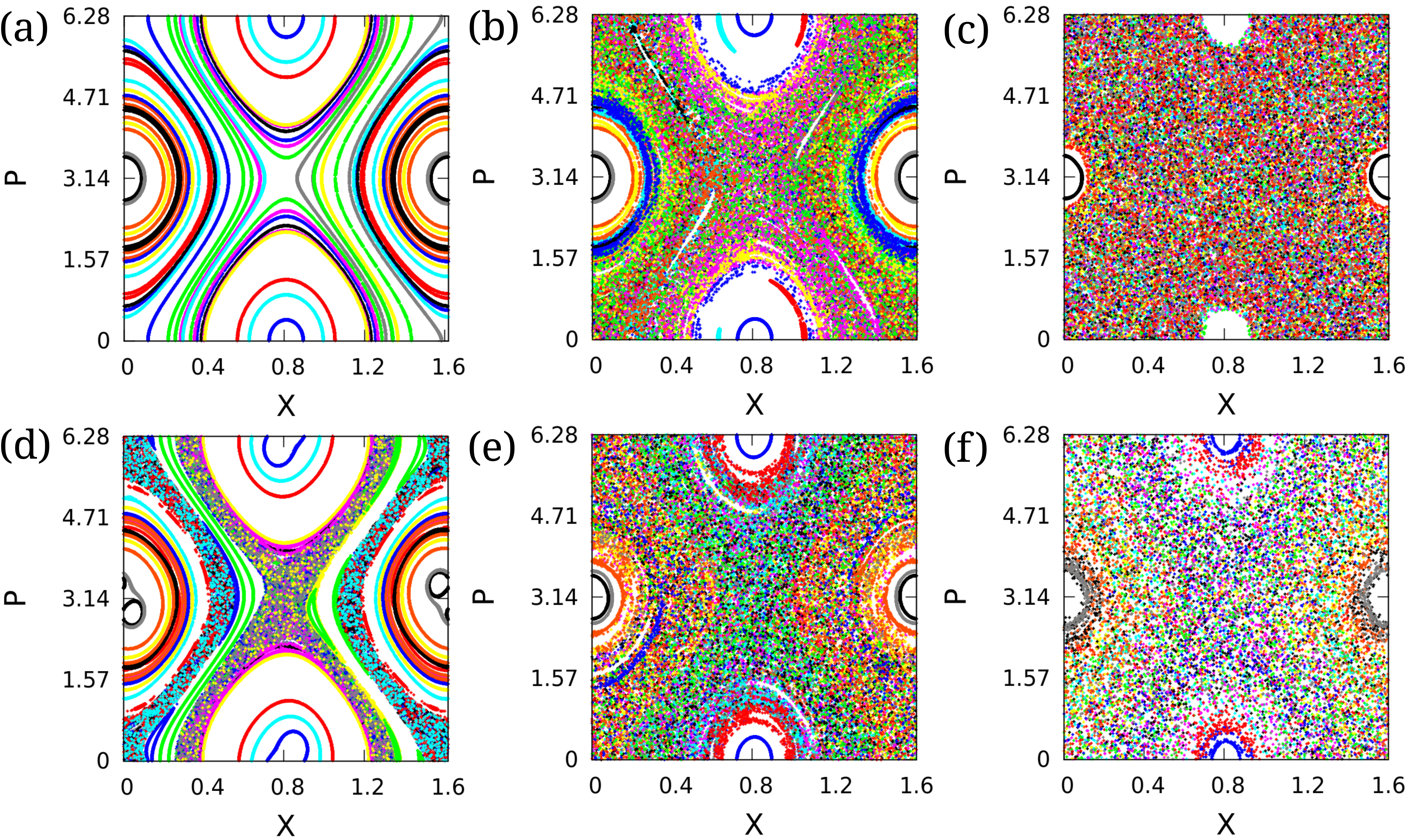}
\caption{Phase space trajectories with increasing values of $\epsilon = 0, 0.033, 0.33$ for $T = 10$ (a-c) and for increasing values of $T = 1, 10, 50$ for $\epsilon = 0.1$ (d-f). Here and in all other figures we set $\lambda = 3$. Different colors indicate different initial conditions.}
\label{xp_dynamics}
\end{figure}

By evolving the equation of motion given in Eq.\ \ref{eq_of_motion}, we obtain the phase space dynamics in the $X-P$ plane. To analyze the dynamical behavior of the system we plot the classical variables stroboscopically by tuning the strength, $\epsilon$, and time period of the periodic modulation, $T$. To this end we fix the strength of the quasiperiodic potential at $\lambda > 2$ such that it corresponds to the localized regime of the AA model. For sufficiently small value of $\epsilon$ the regular orbits remain stable under periodic perturbation; increase in $\epsilon$ leads to the breaking of such periodic orbits and the system gradually enters into the chaotic regime (see Fig.\ \ref{xp_dynamics}). To elucidate the effect of the frequency of the drive we fix the strength of the periodic perturbation $(\epsilon)$ and observe that the system crosses over from regular to chaotic phase space dynamics as shown in Fig.\ \ref{xp_dynamics} with increasing time period T of the drive. The dynamical instability can also be assessed by a Floquet matrix and analyzing its eigenvalues as is illustrated in details in \cite{supp}.
  
\begin{figure}[ht]
\centering
\includegraphics[scale=0.165]{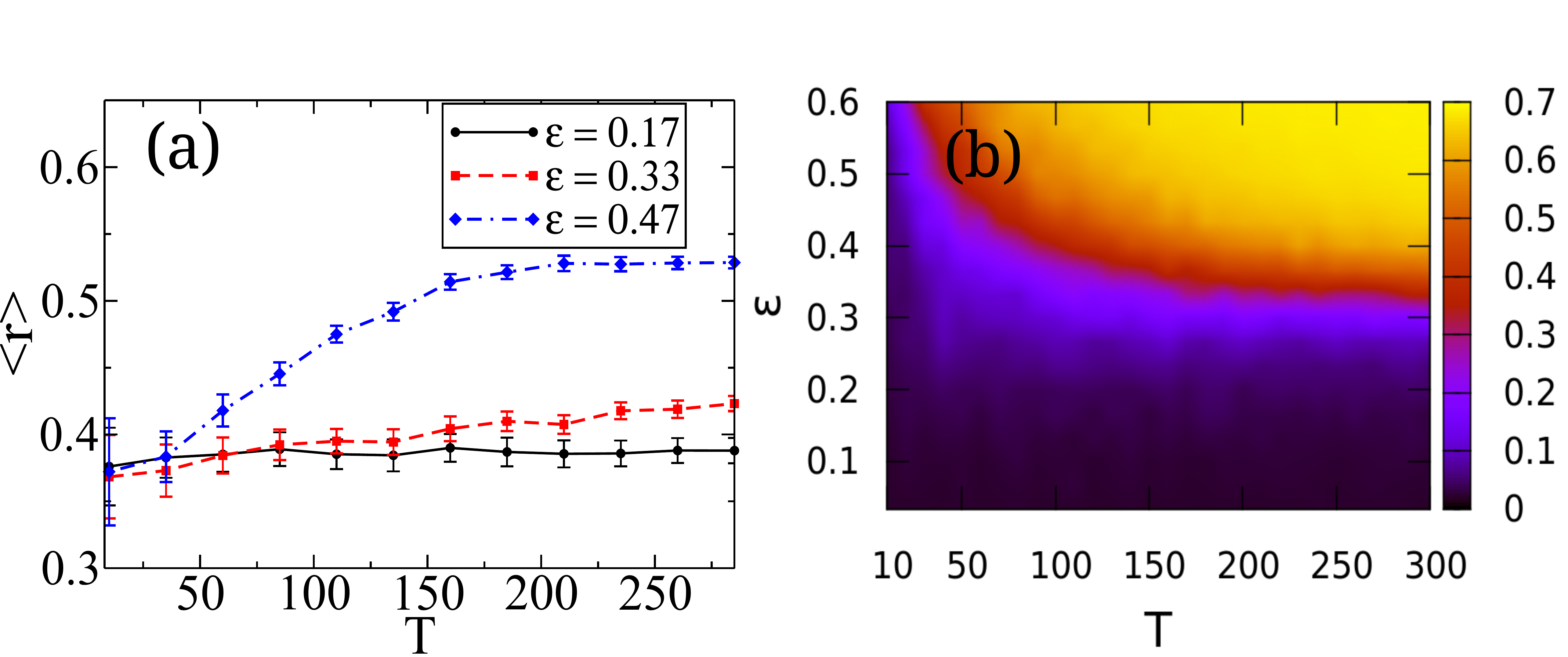}
\caption{(a) $\langle r \rangle$ as a function of $T$ for different values of the driving strength $\epsilon$. (b) Phase diagram in the $T-\epsilon$ plane where color bar indicates $S_{en}$. The delocalization phase boundary asymptotically converges to $\epsilon_c = 1/3$.}
\label{PD_ravg}
\end{figure}

Now we focus on the original quantum Hamiltonian in Eq.\ \ref{ham1} and study the spectral properties of the corresponding Floquet operator $\mathcal{\hat{F}}$. The eigensystem of $\mathcal{\hat{F}}$ are obtained by numerical diagonalization and the eigenphases $\phi_{\nu}$ are ordered in $[-\pi,\pi]$. 
To quantify the statistics from the Floquet spectrum, we calculate the ratio between the consecutive level spacing $r_{\nu}$ given by,
\begin{equation}
r_{\nu} = \frac{\text{min}(\delta _{\nu+1},\delta _{\nu})}{\text{max}(\delta _{\nu+1},\delta _{\nu})}
\end{equation}
where $\delta _{\nu} = \phi_{\nu + 1} - \phi_{\nu}$. We compute the average level spacing ratio $\langle r \rangle$ to identify the degree of delocalization/thermalization. 
In the localized regime, $\langle r \rangle = 2 \ln 2 - 1 \approx 0.386$ signifying that the normalized spacing distribution follows Poisson statistics, whereas in the delocalized state $\langle r \rangle \approx 0.527$ corresponds to the orthogonal class of RMT \cite{lev_stat,bogomolny}. 
For sufficiently large strength $\epsilon$ the value of $\langle r \rangle$ gradually increases from $0.386$ and reaches a value $0.527$ with the increase in the time period, $T$, (see Fig.\ \ref{PD_ravg}a) indicating a drive induced thermalization. 
We further calculate the inverse participation ratio (IPR) of the Floquet eigenstates which is a measure of localization and is given by, $I_{\nu} = \sum_l |\psi_{\nu}(l)|^4$. with increase in $T$, $I_{\nu}$ decreases showing the delocalization of the Floquet eigenstates in the real space \cite{supp}.
We observe that the drive induced delocalization of the Floquet states is connected with the dynamical instability and the onset of chaos found in the underlying classical dynamics. 
It is important to note that our analysis based on the single particle Hamiltonian qualitatively captures the experimentally observed scenarios \cite{bloch}. 

To supplement the connection between the delocalization and the dynamical chaos we further calculate the single particle entanglement entropy (SPEE) of the Floquet eigenstates by dividing the system into two equal parts namely A and B. The reduced density matrix for the subsystem A, corresponding to the $\nu$th Floquet state is given by $\hat{\rho}_{A} = Tr_B |\psi_{\nu}\rangle \langle \psi_{\nu}|$. It can be shown that $\hat{\rho}_{A}$ has two non-zero eigenvalues, $\lambda_1 = \sum_{l\in A} |\psi_{\nu}(l)|^2$ and $\lambda_2 = 1 - \sum_{l\in A} |\psi_{\nu}(l)|^2$ \cite{fradkin}. Thus the SPEE corresponding to the $\nu$th Floquet state can be written as,
\begin{equation}
S^{\nu}_{en} = -\lambda_1 \log \lambda_1 - \lambda_2 \log \lambda_2 
\end{equation}
which vanishes for single site localized states and attains the maximum value $\log2$ for the delocalized states. For large enough driving strength $\epsilon$, we find that the distribution of SPEE of the Floquet eigenstates is peaked around zero for very small time period $T$. On the other hand, for large $T$ the peak of the distribution of SPEE of the Floquet eigenstates is shifted towards the value $\log 2$ \cite{supp} signifying the delocalization of the single particle Floquet states. As a result we observe that the average SPEE $S_{en} = \sum_{\nu = 1}^{\mathcal{N}}S^{\nu}_{en}/\mathcal{N}$ increases from zero with increasing $T$ and reaches the value $\log 2$ for large $T$ as shown in Fig.\ \ref{PD_ravg}(b).        
The effect of the drive parameters $\epsilon$ and $T$ on the delocalization of the Floquet states can be understood from the variation of both $\langle r \rangle$ and $S_{en}$ as summarized in Fig.\ \ref{PD_ravg}.
We find that this delocalization process requires a driving strength above the threshold value of $\epsilon_c = 1-2/\lambda$ for which there is an adiabatic mixing between the localized and the delocalized states of the AA model. 

\begin{figure}[ht]
\centering
\includegraphics[scale=0.12]{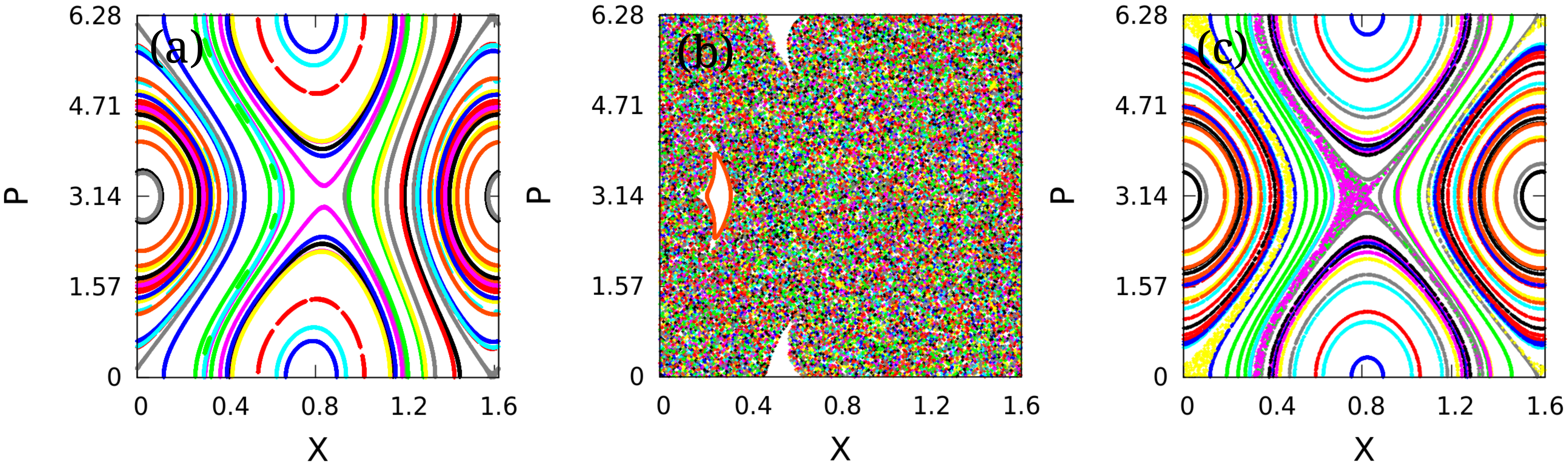}
\caption{(a-c) Phase space trajectories for increasing values of $T = 0.1, 0.7, 10$. Here and in rest of the plots we set $\Delta = 1$.}
\label{PP2}
\end{figure}

\begin{figure}[ht]
\centering
\includegraphics[scale=0.135]{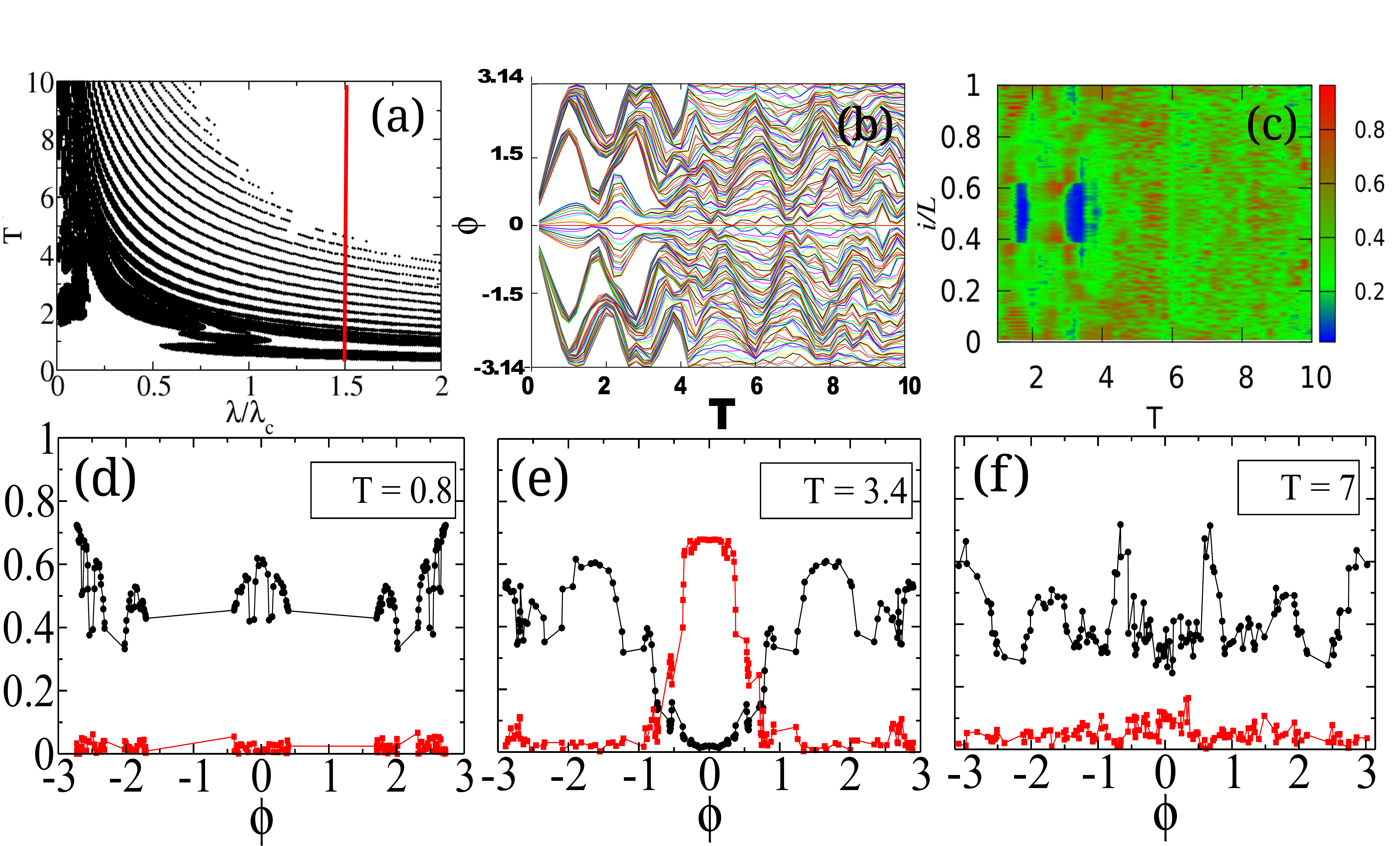}
\caption{(a) Stability of $\bar{R}$ in the $T - \lambda/\lambda_c$ plane. (b) Floquet eigenspectrum. (c) Color plot of IPR for different Floquet states. (d-f) IPR of all the eigenmodes are plotted for different values of $T$.}
\label{fig5}
\end{figure}

{\it Model II:}
Unlike the first model here we consider a periodic drive generated by applying a time dependent gauge field $A(t)$ which in turn gives rise to a periodic modulation of the hopping amplitude $J_{i, j} = J e^{\imath \int_{i}^{j}A(t)dx}$.
The present system is described by the Hamiltonian,
\begin{subequations}
\begin{eqnarray}
\hat{H}_{0} &=& \lambda \sum_{l}\cos(2\pi \beta l)\hat{n}_{l} \\
\hat{H}_{1}(t) &=& -J\sum_{l} \left(\hat{b}_{l}^{\dagger}\hat{b}_{l+1}e^{\imath \Delta f(\omega t)}+h.c.\right)
\end{eqnarray}
\label{ham2}
\end{subequations}
where $A(t)=\Delta f(\omega t)/a$ is the applied time periodic gauge field of frequency $\omega$. 
We consider a smooth time periodic drive $f(\tau) = \sin \tau$. 
Above model can be described in continuum by the Hamiltonian,
\begin{equation}
H(X,P) = -2\cos(P + \Delta \sin \omega t) + \lambda \cos (2\pi \beta X)
\end{equation}
where $X$ and $P$ are the dimensionless canonical variables described earlier. The corresponding Hamilton's equation of motion is given by,
\begin{subequations}
\begin{eqnarray}
\dot{X} &=& 2\sin(P + \Delta \sin \omega t) \\
\dot{P} &=& 2\pi \beta \lambda \sin (2\pi \beta X)
\end{eqnarray} 
\label{eq_of_motion2}
\end{subequations}
We present our analysis for a fixed value of coupling $\lambda =3$ corresponding to the localized regime of the AA-model.  
Using Eq.\ \ref{eq_of_motion2}, we study the stroboscopic dynamics for increasing period of the drive $T$ and keeping the strength fixed at $\Delta =1$ as depicted in Fig.\ \ref{PP2}. For small $T$, regular periodic orbits are formed which is also expected since the time averaging at high frequency gives rise to AA-model with reduced effective hopping amplitude $J {\mathcal J}_{0}(\Delta)$, where ${\mathcal J}_{0}$ is Bessel function of order zero \cite{dalibard_prx,rafael}. This scenario is observed in Fig.\ \ref{PP2}(a) corresponding to the localization. For an intermediate value of the time period $T$ the onset of chaos is identified from the mixed phase-space of the stroboscopic plots. Within this window of time period, fine tuning of $T$ results in fully chaotic phase-space as shown in Fig.\ \ref{PP2}(b).  
With increase in $T$, stable islands re-appear again in the stroboscopic plot while the chaotic regions shrink and finally periodic orbits occupy most of the phase space indicating regular dynamics for large $T$ [see Fig.\ \ref{PP2}(c)]. Appearance of chaotic dynamics within an intermediate range of the time period $T$ is a non trivial effect induced by the time periodic gauge field which has interesting consequences. We also find that the width of $T$ for onset of chaos, increases with the amplitude $\Delta$ of the drive.
Since there is no fixed point of Eq.\ \ref{eq_of_motion2}, we consider the classical Floquet dynamics around the phase-space point $\bar{R}$ which is the stable fixed point in absence of drive at $t=0$. We numerically construct the Floquet matrix $F$ corresponding to $\bar{R}$ to perform the stability analysis as outlined in details in \cite{supp}.
The region of instability in $T-\lambda$ plane is shown in Fig.\ \ref{fig5} which qualitatively explains the onset of chaos and its dependence on $T$. 
Both the time period $T$ and its width corresponding to the instability region decreases with increasing $\lambda$ indicating enhanced stability of periodic motion in more localized regime.

Next we analyze the quantum counterpart and obtain the eigenspectrum of the corresponding Floquet operator constructed from the Hamiltonian in Eq.\ \ref{ham2}. The eigenphases ordered between $[-\pi,\pi]$ are shown in Fig.\ \ref{fig5}(b) which reveals interesting features for increasing time period $T$. In the region of small T the appearance of large energy gaps is reflected in the staircase like structure (see in \cite{supp}) in the integrated density of states (IDOS) which resembles the fractal like structure in AA model \cite{sray_njp}. Whereas with increasing T the large gaps in the quasi-energy spectrum are reduced giving rise to nearly continuous IDOS. To investigate the delocalization of the quasienergy states we compute both IPR and single particle EE as outlined in Model I. As seen in Fig.\ \ref{fig5}(c) we identify an interval of T where islands of delocalized states appear. In Fig.\ \ref{fig5}(e) we closely analyze the spectrum for a typical T within this interval and observe the formation of delocalized Floquet band at the center separated from the localized states by a mobility edge. 
We emphasize that the dynamical instability and onset of chaos within the interval of T (as indicated by a straight line in Fig.\ \ref{fig5}(a)) leads to the formation of such islands of delocalized bands.  
Both above and below this interval of T all the Floquet states are localized [see Fig.\ \ref{fig5}(d,f)] in accordance with the dynamical stability.

\begin{figure}[ht]
\centering
\includegraphics[scale=0.48]{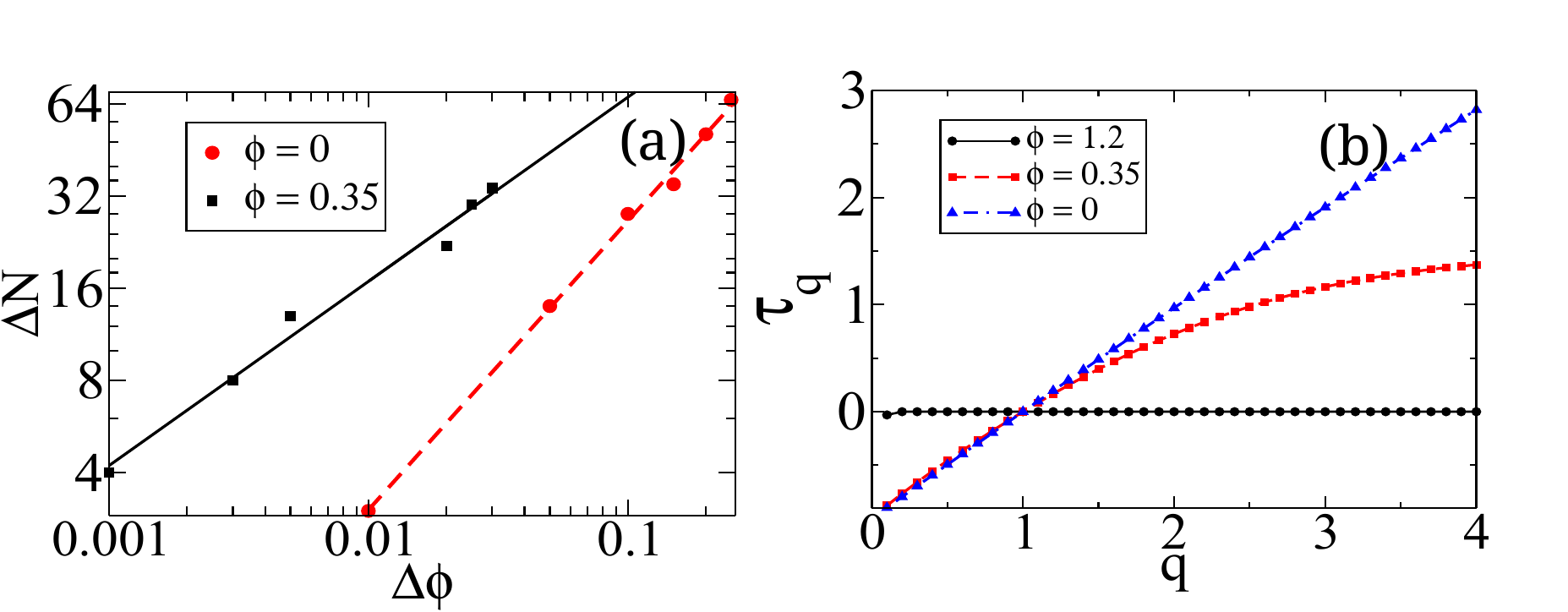}
\caption{(a) Scaling of $\Delta N$ with $\Delta \phi$ with different exponents for band center ($\phi \sim 0$) and band edge ($\phi \sim 0.35$). (b) $\tau_q$ vs $q$ for different eigenphase at $T = 3.4$.}
\label{fig6}
\end{figure}

The existence of the staircase like structure in the spectrum together with the mobility edge observed in the delocalized band motivate us to explore the possible multifractal nature of the spectrum \cite{frac_spec,tgeisel}. 
In general the local number of states $\Delta N$ within an interval $\Delta \phi$ around the eigenphase $\phi$ follows the scaling law, $\Delta N \sim (\Delta \phi)^{\alpha(\phi)}$; dependence of $\alpha$ on the eigenphase $\phi$ signifies the multifractality in the spectrum \cite{pandit,kohmoto,tgeisel}.
For the delocalized band corresponding to Fig.\ \ref{fig5}(e) we find that the exponent $\alpha \sim 0.95$ at the center of the band, whereas, it changes to $\alpha \sim 0.6$ near the mobility edge as illustrated in Fig.\ \ref{fig6}(a).
We also investigate the multifractality of the corresponding Floquet eigenstates from the scaling of their moments given by, $I_q = \sum_n |\psi(n)|^{2q} \sim L^{-\tau _q}$ \cite{frac_spec}, where $\psi(n)$ is the amplitude of the Floquet eigenstate $|\psi \rangle$ at $n$th lattice site and $L$ is the total number of lattice sites.
Variation of $\tau_q$ as a function of $q$ is shown in Fig.\ \ref{fig6}(b) for three different regimes of the eigenspectrum.
We observe that near the center of the delocalized band $\tau _q$ behaves linearly with $q$. Whereas near the mobility edge $\tau _q$ shows a non-trivial dependence on $q$ leading to the conclusion that not only the spectrum is multifractal, also the Floquet eigenstates near the mobility edge exhibit multifractality. 
On the other hand, for the localized states $I_q$ becomes independent of $L$ resulting in $\tau_q \sim 0$.
The dynamical signature of the existence of such multifractal states can be tested from the spreading of the wavepacket in cold atom experiments.  

\begin{figure}[ht]
\centering
\includegraphics[scale=0.15]{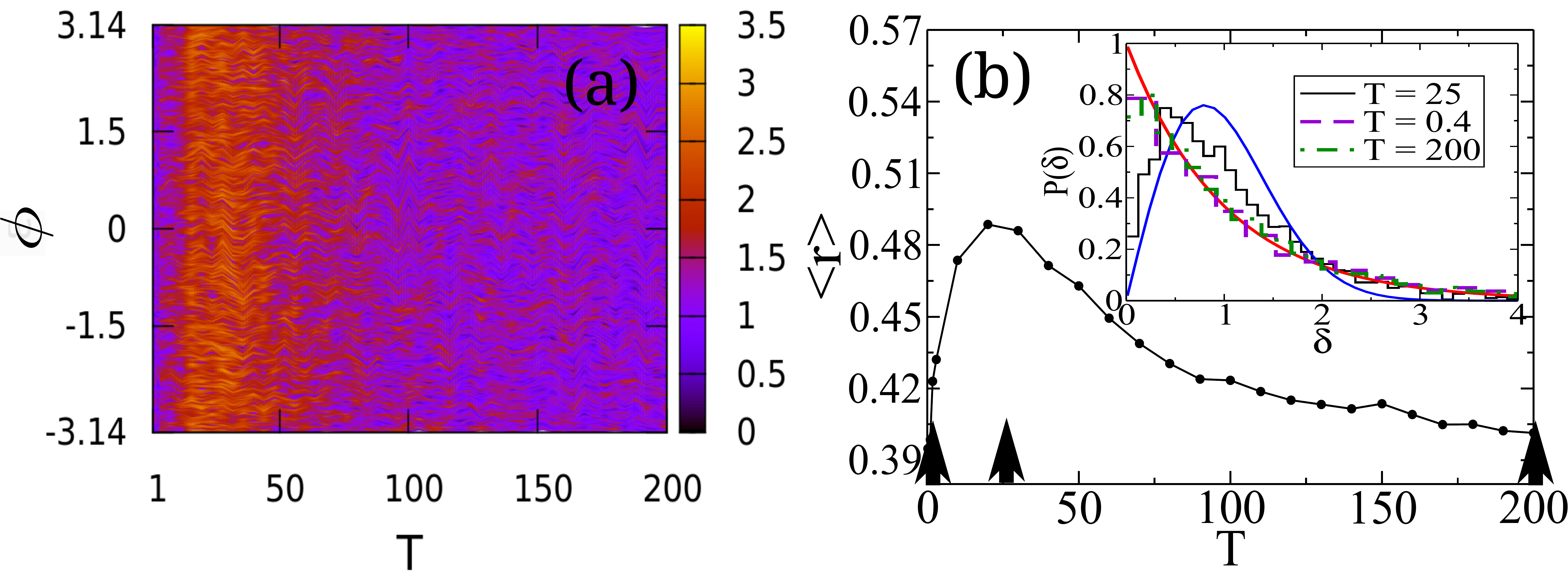}
\caption{(a) Entanglement entropy calculated from the many body eigenstates of the Floquet operator as a function of $T$. (b) $\langle r\rangle$ is plotted as a function of $T$. The spacing distribution of the eigenphases are shown for different values of $T$ (indicated by the arrowhead) in the inset. Other parameters are $\lambda = 3$, $V = 0.1$ and $\Delta = 1$ for system size $L = 14$.}
\label{IPR_int}
\end{figure}

{\it Interaction:} We further investigate the effect of interaction by considering a system of hardcore bosons at half filling with interaction term $H_I = \sum _{l} V \hat{n}_l \hat{n}_{l+1}$ where $V$ is the strength of the nearest neighbor interaction. The time dependent gauge field $A(t)$ can equivalently be described by an electric field $E = - \frac{\partial A}{\partial t} = 4\Delta / T f(\omega t)$ which gives rise to the driving term $H_1(t) = 4\Delta / T f(\omega t) \sum_{l} l \hat{n}_l$ in the Hamiltonian \cite{arimondo}. For simplicity we consider a square wave drive $f(x) = \theta(x-\pi/2) - 2\theta(x-3\pi/2) - \theta(\pi/2-x)$. We calculate the eigensystem of the corresponding Floquet operator using the exact diagonalization method. 
We calculate the entanglement entropy $S^{\nu}_{ent}$ from the many body eigenstates $|\nu\rangle$ by dividing the lattice into two equal parts and shown by color scale plot in Fig.\ \ref{IPR_int}(a). From the relative variation of $S^{\nu}_{ent}$, the appearance of the more delocalized states is clearly visible within an interval of $T$. 
Although this non-monotonic behavior is indeed in agreement with that obtained from the single particle analysis, however we notice that the mobility edge disappears in the presence of interaction.
We further calculate the average level spacing ratios $\langle r\rangle$ and have plotted it as a function of $T$ in Fig.\ \ref{IPR_int}(b). In the limit of small as well as large $T$, $\langle r\rangle \sim 0.386$ signifying the localized Floquet states in the two extremes; the corresponding spacing distribution of the eigenphases exhibits Poisson distribution as shown in the inset of Fig.\ \ref{IPR_int}(b). In the intermediate regime, $\langle r\rangle$ increases from $0.386$ with increasing $T$ and shows a peak around the region of the delocalized Floquet states signifying the existence of level repulsion in the eigenspectrum [see the inset of Fig.\ \ref{IPR_int}(b)].      

{\it Conclusion:} To summarize, in this work we studied single particle delocalization phenomena in  periodically driven AA model and propose a classical route leading to such process.  In {\it Model I} the quasi-periodic potential is periodically modulated, similar to the experimental setup, which leads to delocalization by increasing the time period of the drive \cite{bloch}. Our analysis suggests that the essential features of the experimentally observed drive induced thermalization from the MBL phase can be explained from the single particle picture. Moreover, we find apart from the frequency the amplitude of the drive is equally important and a minimum threshold is required for delocalization which can be tested experimentally. 
In {\it Model II} we consider the AA model with modulated hopping amplitude by the application of a time periodic gauge field which gives rise to a nontrivial effect. Within a certain range of time period of the drive, the delocalized Floquet band with mobility edge appears. Near the mobility edge both the spectrum and eigenfunctions exhibit multifractality. 
The dynamical signature of the existence of such multifractal states can be investigated from the spreading
of the matter wave in cold atom experiments by periodically shaking the bichromatic optical lattice \cite{arimondo}.
In the presence of interaction we found that the non-monotonic dependence of the delocalization on driving period is in accordance with the single particle analysis, however, the mobility edge disappears.
The observed signature of level repulsion in the delocalizad band and its relation to thermalization is worthwhile to explore experimentally by monitoring the decay of the imbalance factor \cite{bloch_mbl}.

\newpage

\begin{widetext}
\appendix

\begin{center}
{\bf SUPPLEMENTAL MATERIAL}
\end{center} 
 
In this supplemental material we provide the details of the stability analysis of the classical dynamical equations. We also discuss the behavior of inverse participation ratio (IPR) and the single particle entanglement entropy (SPEE) of the Floquet states corresponding to the models discussed in the main text.   

\section{Stability Analysis}

To perform a stability analysis around a point in the phase space constructed from the dynamical variables $X$ and $P$, we evolve an initial point $R^0 \equiv (X_0,P_0)$ upto one time period $T$. So, the final state after time $T$ can be written as a map $R^T = F R^0$ \cite{arnold,pedersen}, where $F$ is the Jacobian matrix which governs the evolution of the dynamical variables for a time period $T$. We numerically construct $F$ whose elements are given by, $F_{ij} \equiv \partial R_i^T/\partial R_j^0$. The instability of $R^0$ sets in when the eigenvalues of $F$ satisfies $|\Lambda| > 1$.     

The dynamical equations (see Eq.\ [5] in main text) has a pair of fixed points $\bar{R} \equiv \{(0,\pi), (1/2\beta,0)\}$ whose stability is shown by tuning the parameters $\lambda$ and $T$ in Fig.\ \ref{PD_stability_supp}. We see that there are stripes of black bands where the fixed points $\bar{R}$ become unstable. With increasing $T$ these regions of instability grow and finally the phase space becomes completely chaotic. 
The dynamical equations (see Eq.\ [5] in main text) can be further linearized around a fixed point, say ($0,\pi$) and are given by,
\begin{subequations}
\begin{eqnarray}
\delta \dot{X} &=& -2\delta P \\
\delta \dot{P} &=& \lambda (2\pi \beta)^2 (1 + \epsilon f(\omega t))\delta X
\end{eqnarray}
\label{lin_eqn}
\end{subequations}
Thus the Jacobian matrix governing the time evolution of $\delta X$ and $\delta P$ within the time period $T$ can be expressed as,
\begin{equation}
F = F_1F_2, \quad F_i = \left( \begin{array}{cc}
\cos \omega _i T/2 & -(4/\omega _i) \sin \omega _i T/2 \\
(\omega _i/4) \sin \omega _i T/2 & \cos \omega _i T/2 \end{array}\right)
\end{equation}
where, $\omega _{1(2)} = 2\pi \beta \sqrt{2\lambda} \sqrt{1 +(-) \epsilon}$. The system evolves from $0$ to $T/2$ under $F_1$ and from $T/2$ to $T$ under $F_2$. The instability condition leads to a simpler expression given by,
\begin{equation}
2\cos (\omega_1 T/2) \cos (\omega_2 T/2) - (\omega_1/\omega_2 + \omega_2/\omega_1) \sin (\omega_1 T/2) \sin (\omega_2 T/2) > 2 
\end{equation}
The region of instability obtained using the above expression agrees well with our numerical analysis. 

\begin{figure}[ht]
\centering
\includegraphics[scale=0.25]{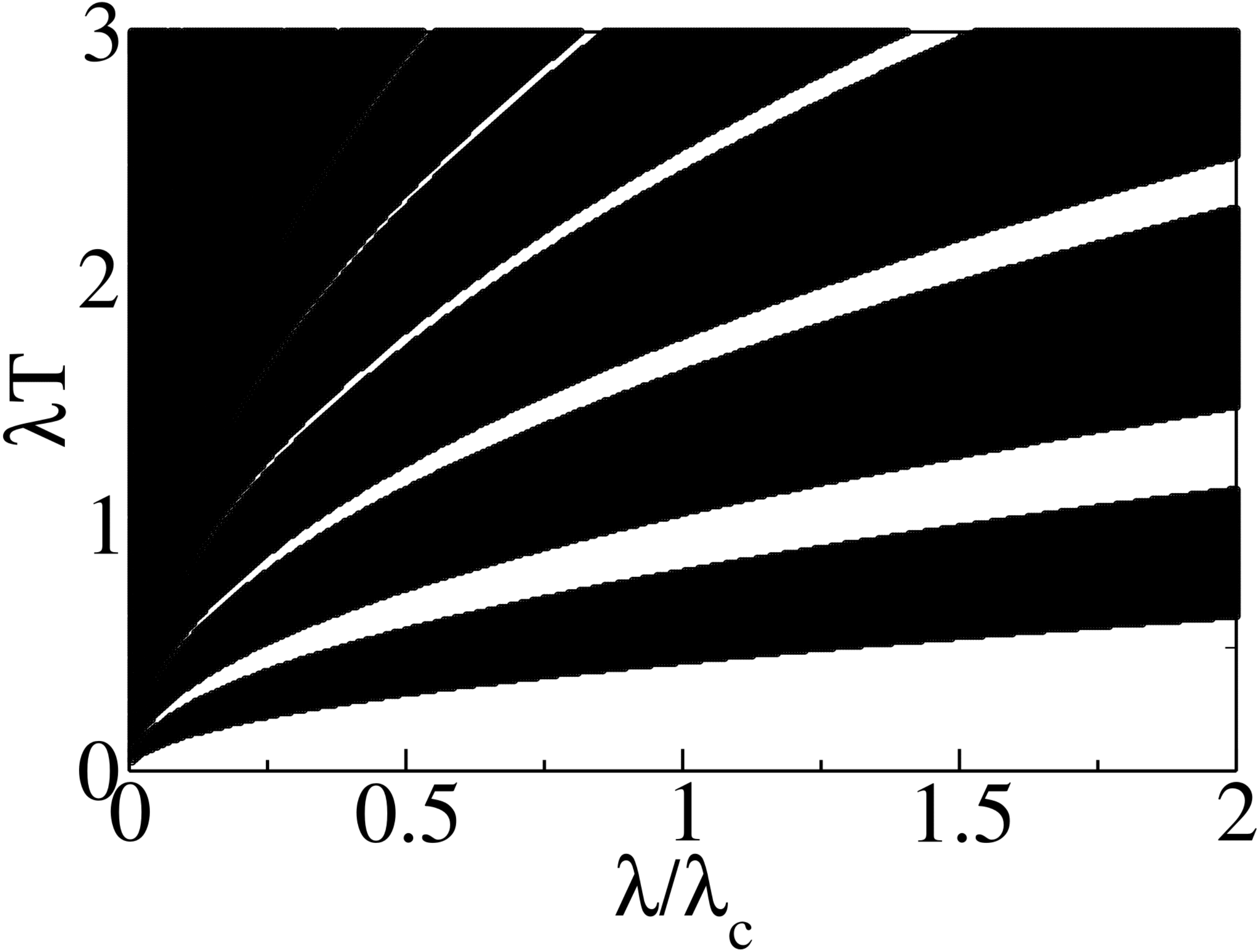}
\caption{A stability diagram has been shown in $\lambda T-\lambda/\lambda_c$ plane for $\epsilon = 1$. See the text for the details.}
\label{PD_stability_supp}
\end{figure}  

\section{Spectral properties of the Floquet matrix}

\subsection{Model I}

From the spectral analysis of $\mathcal{\hat{F}}$ we found that the average spacing ratios take the value $0.386$ which corresponds to Poisson statistics for smaller $T$ and with increasing $T$ it gradually rises to $0.527$ signifying the orthogonal class of the random matrix theory (RMT). Here we compute the normalized spacing distribution of the Floquet eigenphases for the two extreme regimes of $T$ and are illustrated in Fig.\ \ref{lev_stat}(a-b). 

\begin{figure}[ht]
\centering
\includegraphics[scale=0.17]{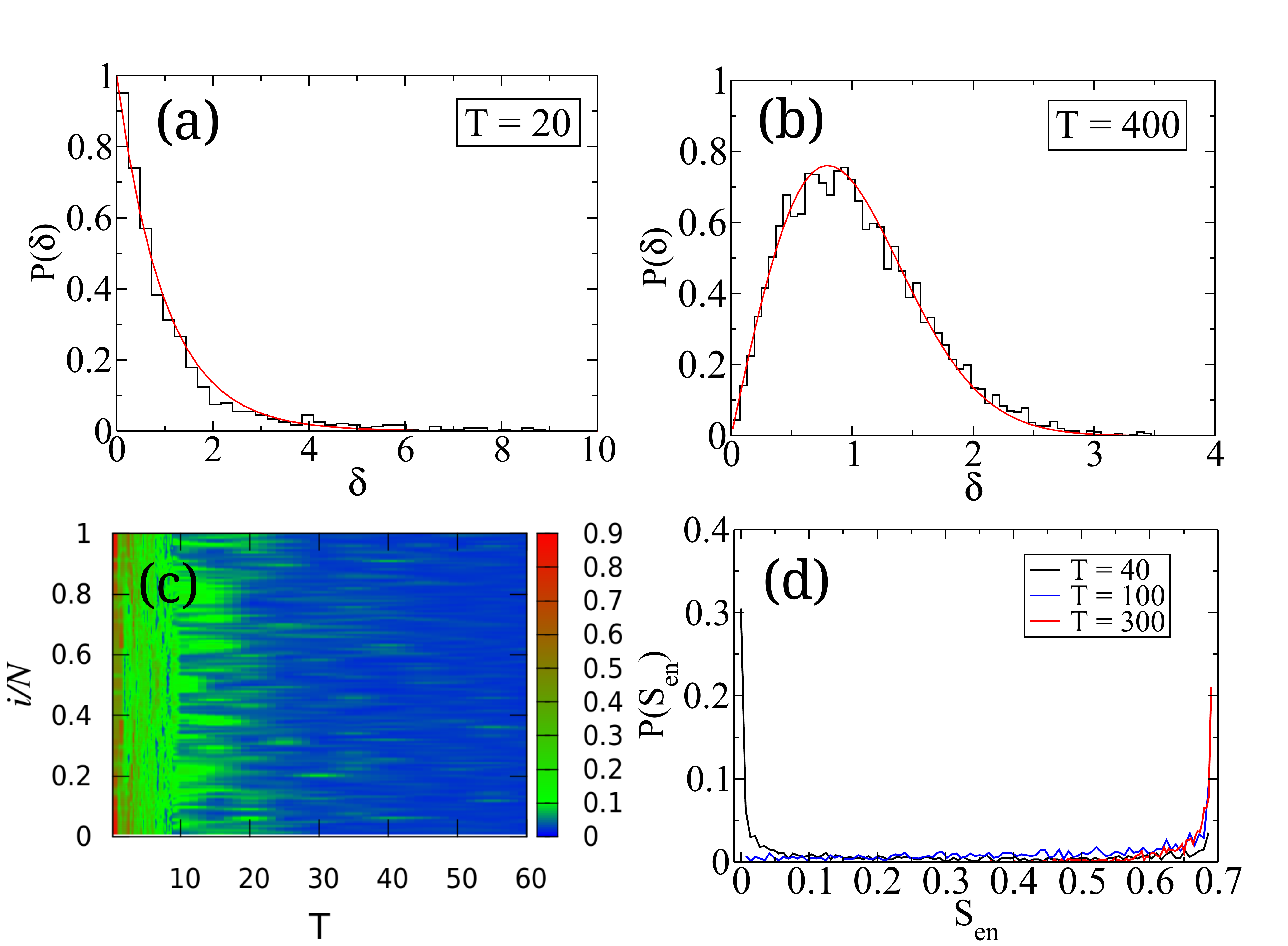}
\caption{(a-b) Spacing distribution of the eigenphases is shown for different values of $T$ mentioned in the inset. For smaller driving time period $T$, the distribution follows Poisson statistics, whereas for higher $T$ the distribution follows Wigner-Surmise; the corresponding probability distributions are shown by a solid curve (red). (c) IPR for different Floquet states are shown as a function of $T$. (d) Distributions of the SPEE are plotted for different values of $T$ mentioned in the inset. Other parameters are $\lambda = 3$ and $\epsilon = 0.6$.}
\label{lev_stat}
\end{figure}  

To elucidate the delocalization of the Floquet states with increasing $T$ in the regime for large $\epsilon$, we compute the inverse participation ratio (IPR) and shown as a function of $T$ in Fig.\ \ref{lev_stat}(c). Further we calculate the distributions of the SPEE for different regimes of the driving time period $T$ and are shown in Fig.\ \ref{lev_stat}(d). For smaller value of $T$, the distribution is peaked near zero signifying the localized phase, whereas, for large $T$ the peak of the distribution is shifted to $\sim \log 2$ denoting the delocalization of the Floquet states. In the intermediate regime the Floquet states show a flat distribution of the entanglement entropy.    

\subsection{Model II}

{\it Effective Hamiltonian:} The Hamiltonian in the presence a time dependent gauge field is written as,
\begin{equation}
H(X,P,t) = -2\cos(P + \Delta \sin \omega t) + \lambda \cos(2\pi \beta X)
\end{equation} 
Using the identity $e^{\imath \Delta \sin \omega t} = \sum_l \mathcal{J}_l(\Delta) e^{\imath l\omega t}$, where $\mathcal{J}_l$ is the Bessel function of order $l$, we can write the above equation as, 
\begin{equation}
H(X,P,t) = -2 \mathcal{J}_0(\Delta)\cos P + \lambda \cos (2\pi \beta X) -2 \sum_{l=1}^{\infty} \mathcal{J}_l(\Delta)[\cos(P+l\omega t) + (-1)^l\cos(P-l\omega t)]
\end{equation}
Therefore, the time-averaged Hamiltonian is given by $H_{av} = 1/T\int_0^TH(X,P,t)dt$ \cite{dalibard,rafael_supp}, which can be written within the zeroth order approximation as,
\begin{equation}
H_{av} = -2 \mathcal{J}_0(\Delta)\cos P + \lambda \cos (2\pi \beta X)
\end{equation}
$\mathcal{J}_o$ is the zeroth order Bessel function.

{\it Analysis of the Floquet eigensystem:} The eigenphases of the Floquet operator constructed from the Hamiltonian in Eq.\ [8] are ordered between $[-\pi,\pi]$. The eigenspectrum as a function of the driving time period $T$ is shown in Fig.\ 4b and reveals interesting feature. In the smaller $T$ regime the spectrum shows large band gaps whereas with increasing $T$ these large band gaps vanish. To further analyze this observation we calculate the integrated density of states (IDOS) and plotted it as a function of rescaled eigenphase $\phi \equiv (\phi - \phi_{min}) / (\phi _{max} - \phi_{min})$ in Fig.\ \ref{den_state}(a). We see that in the lower $T$ regime the IDOS shows a staircase like structure where the plateaus signifies the large band gaps present in the spectrum. On the other hand for increasing $T$ the vanishing of band gaps result in a nearly continuous IDOS.

\begin{figure}[ht]
\centering
\includegraphics[scale=0.2]{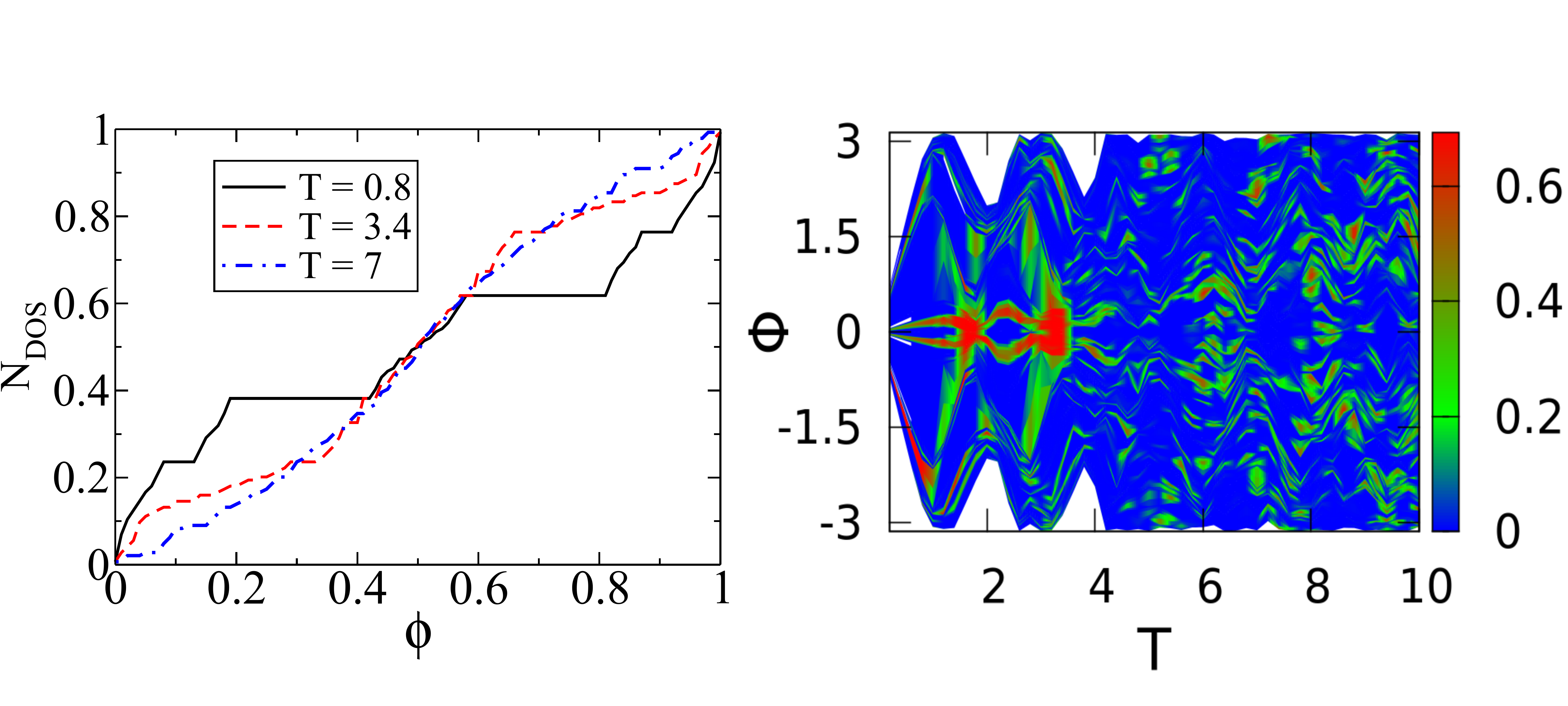}
\caption{(a) Normalized IDOS is plotted for three different values of $T$ indicated in the inset. (b) Single particle EE for the Floquet states are plotted as a function of $T$. Other parameters are $\lambda = 3$ and $\Delta = 1$.}
\label{den_state}
\end{figure}  

Next we calculate the single particle EE (see Eq.\ [7] in main text) for the Floquet states. In Fig.\ \ref{den_state}(b) we have shown the variation of the single particle EE of the Floquet states with increasing $T$. This picture is analogous to the behavior of IPR [see Fig.\ 4(c) in main text] and shows the appearance of the delocalized band characterized by the vanishing IPR and the single particle EE takes the value $\log 2$.

\end{widetext} 
 
\end{document}